\documentclass[final,hidelinks,onefignum,onetabnum]{siamart251216}

\usepackage{lipsum}
\usepackage{amsfonts}
\usepackage{graphicx}
\usepackage{epstopdf}
\usepackage{algorithmic}
\ifpdf
  \DeclareGraphicsExtensions{.eps,.pdf,.png,.jpg}
\else
  \DeclareGraphicsExtensions{.eps}
\fi

\newsiamremark{remark}{Remark}
\newsiamremark{hypothesis}{Hypothesis}
\crefname{hypothesis}{Hypothesis}{Hypotheses}
\newsiamthm{claim}{Claim}
\newsiamremark{fact}{Fact}
\crefname{fact}{Fact}{Facts}

\headers{Time Evolution on Hybrid Tensor Networks}{T. V. Bauer, R. M. Milbradt, and C. B. Mendl}

\usepackage{amsopn}

\usepackage{graphicx} 
\usepackage{xcolor}
\usepackage{mathabx}
\usepackage{braket}
\usepackage[bb=boondox]{mathalfa}
\colorlet{light_grey}{gray!40}
\usepackage{tikz}
\usepackage{tikzit}
\usepackage{orcidlink}
\usetikzlibrary{calc, shapes, positioning, backgrounds, arrows.meta}

\tikzstyle{NONE}=[fill=white, draw=none, circle, minimum size=2pt, inner sep=0pt, tikzit draw=none, tikzit fill={rgb,255: red,128; green,128; blue,128}]
\tikzstyle{medium dist tikzit}=[fill=white, draw=black, shape=regular polygon, tikzit category=kernels, shape border rotate=90, regular polygon sides=3, tikzit shape=regular polygon, minimum height=1cm]
\tikzstyle{medium kernel tikzit}=[fill=white, draw=black, shape=rectangle, tikzit category=kernels, minimum width=.75cm, minimum height=0.75cm]
\tikzstyle{coordinate tikzit}=[tikzit category=coordinates, inner sep=0pt, outer sep=0pt]
\tikzstyle{copymap tikzit}=[fill=black, draw=black, shape=circle, inner sep=1, tikzit category=kernels]
\tikzstyle{small kernel tikzit}=[fill=white, draw=black, shape=rectangle, tikzit category=kernels, minimum width=0.25cm, minimum height=0.25cm]
\tikzstyle{large kernel tikzit}=[fill=white, draw=black, shape=rectangle, tikzit category=kernels, minimum width=1.2cm, minimum height=1.2cm]
\tikzstyle{small circle}=[fill=white, draw=black, shape=regular polygon, tikzit category=kernels, regular polygon sides=3, shape border rotate=90, minimum height=0.5cm, inner sep=-1pt]
\tikzstyle{erase tikzit}=[{-{Rays[n=8]}}]
\tikzstyle{RLabel tikzit}=[anchor=west]
\tikzstyle{Llabel tikzit}=[anchor=east]
\tikzstyle{very small prob}=[fill=white, draw=black, shape=regular polygon, regular polygon sides=3, shape border rotate=90, minimum height=0.5cm, inner sep=-4pt]
\tikzstyle{Circle around node}=[fill=white, draw=blue, shape=circle, inner sep=0pt]
\tikzstyle{plain circle}=[fill=white, draw=black, shape=circle, text width=0.5cm, align=center]
\tikzstyle{LargeCircle}=[fill=white, draw=black, shape=circle, minimum size=1.4cm, align=center]
\tikzstyle{plain yellow}=[fill=yellow, draw=black, shape=circle, text width=0.5cm, align=center]
\tikzstyle{equals node}=[inner sep=0pt, minimum size=0.5cm, draw=none]
\tikzstyle{R_triangle}=[draw, regular polygon, regular polygon sides=3, minimum size=12mm, inner sep=0pt, shape border rotate=30]
\tikzstyle{L_triangle left}=[draw, regular polygon, regular polygon sides=3, minimum size=12mm, inner sep=0pt, shape border rotate=210]
\tikzstyle{black_filled}=[fill=black, draw=black, shape=circle, tikzit fill=black, tikzit draw=black, minimum size=0.5 cm]
\tikzstyle{grey}=[fill={light_grey}, draw=black, shape=circle, tikzit fill=black, tikzit draw=black, minimum size=0.5 cm]
\tikzstyle{red text}=[fill=white, draw=red, shape=circle, text width=0.25cm, align=center, text=red]
\tikzstyle{process}=[fill={rgb,255: red,191; green,191; blue,191}, draw=black, shape=rectangle, tikzit draw=black, tikzit fill={rgb,255: red,191; green,191; blue,191}, text width=2.5cm, align=center]
\tikzstyle{return value}=[fill={rgb,255: red,191; green,191; blue,191}, draw=black, shape=ellipse, tikzit fill={rgb,255: red,191; green,191; blue,191}, tikzit draw=black, tikzit shape=circle]
\tikzstyle{decission}=[fill={rgb,255: red,191; green,191; blue,191}, draw=black, shape=diamond, tikzit fill=yellow, tikzit draw=black, tikzit shape=circle, text width=1cm]
\tikzstyle{quantum process}=[fill={rgb,255: red,225; green,225; blue,225}, draw=black, shape=rectangle, tikzit shape=rectangle, tikzit fill={rgb,255: red,191; green,191; blue,191}, tikzit draw=black, dashed, text width=2.5cm]

\tikzstyle{delmap}=[>={Rays[n=8]}, ->]
\tikzstyle{arrows_double}=[stealth-stealth]
\tikzstyle{red}=[-, draw=red]
\tikzstyle{red dash}=[-, draw=red, dashed, tikzit draw=red]
\tikzstyle{black}=[-, draw=black]
\tikzstyle{blue}=[-, draw=blue]
\tikzstyle{left arrow}=[stealth-]
\tikzstyle{right arrow}=[-stealth]
\tikzstyle{dashed edge}=[-, dashed]
\tikzstyle{dots}=[-, dotted]
\tikzstyle{right bent}=[->, draw=red, tikzit draw=red, bend right=90, line width=1pt]
\tikzstyle{robbit}=[-, draw=black, double=red, double distance=3pt]
\tikzstyle{thick black}=[-, fill=none, draw=black, tikzit draw=black, line width=2pt]
\tikzstyle{dotted blue}=[-, draw=blue, dotted, tikzit draw=blue]
\tikzstyle{process connection}=[-{Stealth[length=3mm]}]
\tikzstyle{process no arrow}=[-]

\def\equationautorefname~#1\null{Eq.~(#1)\null}
\def\figureautorefname~#1\null{Fig.~#1\null}
\def\subfigureautorefname~#1\null{Fig.~#1\null}
\def\sectionautorefname~#1\null{Sec.~#1\null}
\def\tableautorefname~#1\null{Tab.~#1\null}
\def\theoremautorefname~#1\null{Rem.~#1\null}

\usepackage{soul}
\pgfdeclarelayer{edgelayer}
\pgfdeclarelayer{nodelayer}
\pgfsetlayers{background,edgelayer,nodelayer,main}

\title{Time Evolution on Hybrid Tensor Networks - A Novel and Parallelizable Algorithm}
\author{Tobias~Valentin~Bauer\orcidlink{0009-0004-8511-0563}\thanks{Leibniz Supercomputing Centre of the Bavarian Academy of Sciences and Humanities (LRZ), Garching b. München, Germany ({Tobias.Bauer@LRZ.de}).}
\and Richard~M.~Milbradt\orcidlink{0000-0001-8630-9356}\thanks{Technical University of Munich, CIT, Department of Computer Science, Boltzmannstra{\ss}e 3, 85748 Garching, Germany (\email{r.milbradt@tum.de}, \email{christian.mendl@tum.de}).}
\and Christian~B.~Mendl\orcidlink{0000-0002-6386-0230} \footnotemark[2] \thanks{Technical University of Munich, Institute for Advanced Study, Lichtenbergstra{\ss}e 2a, 85748 Garching, Germany.}
}
\date{July 2025}

\begin{document}

\maketitle
\begin{abstract}
We develop a novel time-evolution algorithm for matrix product states based on the recently introduced hybrid tensor network (hTN) framework.
We retain the tensors close to the boundary on the classical computer and offload the highly entangled inner ones to the quantum computer.
In our variant, we employ the Basis Update and Galerkin (BUG) integrator to time-evolve the classical tensors, and we develop a coupling scheme between the classical and quantum parts.
Our framework admits modular combination with any quantum time-evolution method, such as (classically pre-optimized) Trotterization.
The ratio of classical and quantum tensor degrees of freedom can be dynamically adjusted during the time evolution, which can be advantageous when the classical memory requirements become prohibitive. 
The quantum and classical components can run in parallel during a single time step and are not constrained by synchronization barriers or mid-circuit measurements.
We describe the detailed steps and pseudocode for our algorithm specialized for tensor networks originating from the matrix product state Ansatz.
\end{abstract}



\section*{Introduction}

In this work, we present a novel quantum-classical hybrid algorithm for time evolutions. Within one time step, the workload of the classical and quantum computers has no serial data dependency and can run in parallel. 
Furthermore, we describe a subflow for the hybrid tensor networks (hTN) framework that balances additional auxiliary qubits against the number of states to be evolved in the realization of a quantum tensor. In the hTN setting, this means trading off classical indices against quantum indices. 
Another subflow we contribute is converting parts of a Matrix Product State (MPS) into an MPS itself for usage in the hTN context. This enables offloading classical workload to a quantum computer in the context of tensor network algorithms. 
Furthermore, we expand the hTN framework by explicitly defining bookkeeping matrices and their use to move the isometrization center of an MPS over a quantum tensor, which has not been described in detail in previous papers. 

The framework of hybrid tensor networks (hTN) \cite{schuhmacherHybridTreeTensor2024, yuanQuantumSimulationHybrid2021} is an emerging field that combines the strengths of classical tensor network simulation with the power of quantum computing. The work of \cite{yuanQuantumSimulationHybrid2021} introduced the hTN framework and demonstrated the solution of ground-state problems for lattice Hamiltonians using standard imaginary-time evolution schemes. Notably, they used hTNs without a classical component, meaning they use several quantum devices to solve the problem. 
In \cite{schuhmacherHybridTreeTensor2024}, the problem of canonical forms and isometrization for hTN was considered. This allowed them to adapt DMRG to hTN by applying DMRG to the classical parts and performing variational optimization on the quantum tensors in sweeps. During optimization, a sweep cannot continue until the optimization at the previous site has concluded. This creates a situation in which the classical and quantum parts of the algorithm are serially dependent. Just like their main algorithm, variational optimization itself is a hybrid quantum-classical method, with serial dependence of the quantum and classical parts of the algorithm. Our algorithm, however, aims to make use of quantum and classical resources in parallel.

The hTN framework is particularly suited for Matrix Product States (MPS) that can be prepared in constant depth on a quantum computer \cite{smithConstantdepthPreparationMatrix2024}, which holds true for parts of an MPS that are used to initialize the quantum part of the hTN on the quantum computer. We propose an algorithm that uses the thus-prepared hTN to simulate the time evolution of a quantum system. Algorithms performing such a simulation are of relevance, particularly to theoretical chemistry and many-body physics \cite{bauerQuantumAlgorithmsQuantum2020, eisertQuantumManybodySystems2015, caoQuantumChemistryAge2019}. 

However, purely classical time evolutions on a state expressed as a tensor network are bound to fail as the number of timesteps increases due to the entanglement growth. In turn, this increases the memory footprint required to accurately represent the state as a tensor network. Eventually, this representation becomes infeasible, limiting the number of time steps \cite{paeckelTimeevolutionMethodsMatrixproduct2019}. 

Thus, it is reasonable to transfer fully classical tensor network algorithms to hTNs. For the former, the time-evolving block decimation (TEBD) algorithm~\cite{Vidal2003, vidalEfficientSimulationOneDimensional2004} is a standard choice for time-evolving states expressed as tensor networks. However, it suffers from several issues: As noted in \cite{haegemanTimeDependentVariationalPrinciple2011}, TEBD can exhibit non-optimal truncation and potentially violate conservation laws and symmetries. Additionally, it is problematic to implement long-range and many-particle interactions using TEBD \cite{Stoudenmire2010}. The projector-splitting time-dependent variational principle (PS-TDVP) algorithm \cite{haegemanTimeDependentVariationalPrinciple2011, Lubich2015, haegemanUnifyingTimeEvolution2016} is a subsequent advancement in time evolution methods, based on the well-known Dirac-Frenkel variational principle \cite{Lubich2008}. In addition to resolving the aforementioned issues of TEBD, extensions of the PS-TDVP offer the flexibility to dynamically adapt the bond dimension as needed \cite{haegemanUnifyingTimeEvolution2016, Yang2020, Dunnett2021, Xu2022, Li2024, McCulloch2024}. However, this approach requires two time updates per site, which we find to be the most time-consuming part in the context of hybrid tensor networks. Finally, the Basis Update and Galerkin integrator (BUG) \cite{Ceruti2023} is a more recent development in time-evolution algorithms for tensor networks. It merely requires a single forward-in-time update for every tensor, avoiding some instability concerns that can appear for the PS-TDVP. The recently developed parallel variant~\cite{cerutiParallelBasisUpdate2026} is well suited for high-performance computing (HPC) environments.

The algorithm proposed in this paper alleviates the limit of classical hardware by adapting the parallel BUG algorithm to hTNs. The parts of a tensor network with the highest bond dimension are offloaded to quantum hardware. 
The offloading can be done as-needed. Thus, the robustness and speed of classical computers can be fully utilized until memory requirements become prohibitive. The parallel nature of the BUG integrator allows us to minimize expensive communication between quantum and classical resources. Running the quantum and classical parts in parallel also allows us to achieve higher accuracy on the classical part of the tensor network while waiting for the generally slower evaluation on the quantum computer. 


\section{The BUG algorithm}\label{sec:class_bug}
For convenience, we will briefly review the parallel version of the BUG algorithm \cite{cerutiParallelBasisUpdate2026} for time evolution. Even though the algorithm was formulated for general tree tensor networks, we restrict this introduction to the simpler case of a tensor train.
The goal of this method is to solve a differential equation, such as the Schrödinger equation
\begin{equation}
	\frac{d}{dt}\ket{\Psi} = -iH\ket{\Psi},
\end{equation}
where the state is given as an MPS of length $L$
\begin{equation}
	\ket{\Psi} = \sum_{\vec{v} \vec{s}} T^{[1]}_{v_1s_1} T^{[2]}_{v_1s_2v_2} \cdots T^{[L]}_{v_{L-1}s_Lv_L} \ket{s_1, \dots, s_L}
\end{equation}
and $H$ is given as a matrix product operator (MPO)
\begin{equation}
	H = \sum_{\vec{w} \vec{s} \vec{z}} W^{[1]}_{w_1s_1z_1} W^{[2]}_{w_1s_2z_2w_2} \cdots W^{[L]}_{w_{L-1}s_Lz_Lw_L} \ket{z_1, \dots, z_L}\bra{s_1, \dots, s_L}.
\end{equation}
We will now choose the tensor $T^{[r]}$ for $r=\lfloor L / 2 \rfloor$ as the root tensor and assume that the state $\ket{\Psi}$ is in canonical form with regards to $T^{[r]}$. This means
\begin{align}
	\sum_{v_{i-1} s_i} T^{[i]}_{v_{i-1}s_iv_i} \bar{T}^{[i]}_{v_{i-1}s_iv'_i} & = \mathbb{1}_{v_iv'_i} \quad \forall\, i < r \text{ and } \label{eq:left_orthonormal}\\
	\sum_{v_{i} s_i} T^{[i]}_{v_{i-1}s_iv_i} \bar{T}^{[i]}_{v_{i-1}s'_iv_i} & = \mathbb{1}_{v_{i-1}v'_{i-1}} \quad \forall\, i > r. \label{eq:right_orthonormal}
\end{align}
The BUG algorithm is a method for updating the tensors individually using a local Schrödinger equation with an effective Hamiltonian $H_{\text{eff}}^{[i]}$. However, to update a tensor, it must be the center of the canonical form. Assume $T^{[c]}$ is the current canonical center. Let $Q^{[c]}$ be the site-tensor on site $c$ before it became the canonical center. We can move the canonical center to the next tensor on a copy of the MPS, while already updating $T^{[c]}$ by solving the tensor differential equation 
\begin{equation}\label{eq:loc_diff_eq}
	\frac{d}{dt} T^{[c]} = -i H_{\text{eff}}^{[c]} T^{[c]},
\end{equation}
where the effective Hamiltonian is given by 
\begin{equation}
	H_{\text{eff}}^{[i]} = \sum \mathcal{L}^{[i-1]}_{v_{i-1}w_{i-1}\bar{v}_{i-1}} W^{[i]}_{w_{i-1}s_iz_iw_i} \mathcal{R}^{[i+1]}_{v_iw_i\bar{v}_i} \ket{\bar{v}_{i-1}z_i\bar{v}_i} \bra{v_{i-1}s_iv_i}.
\end{equation}

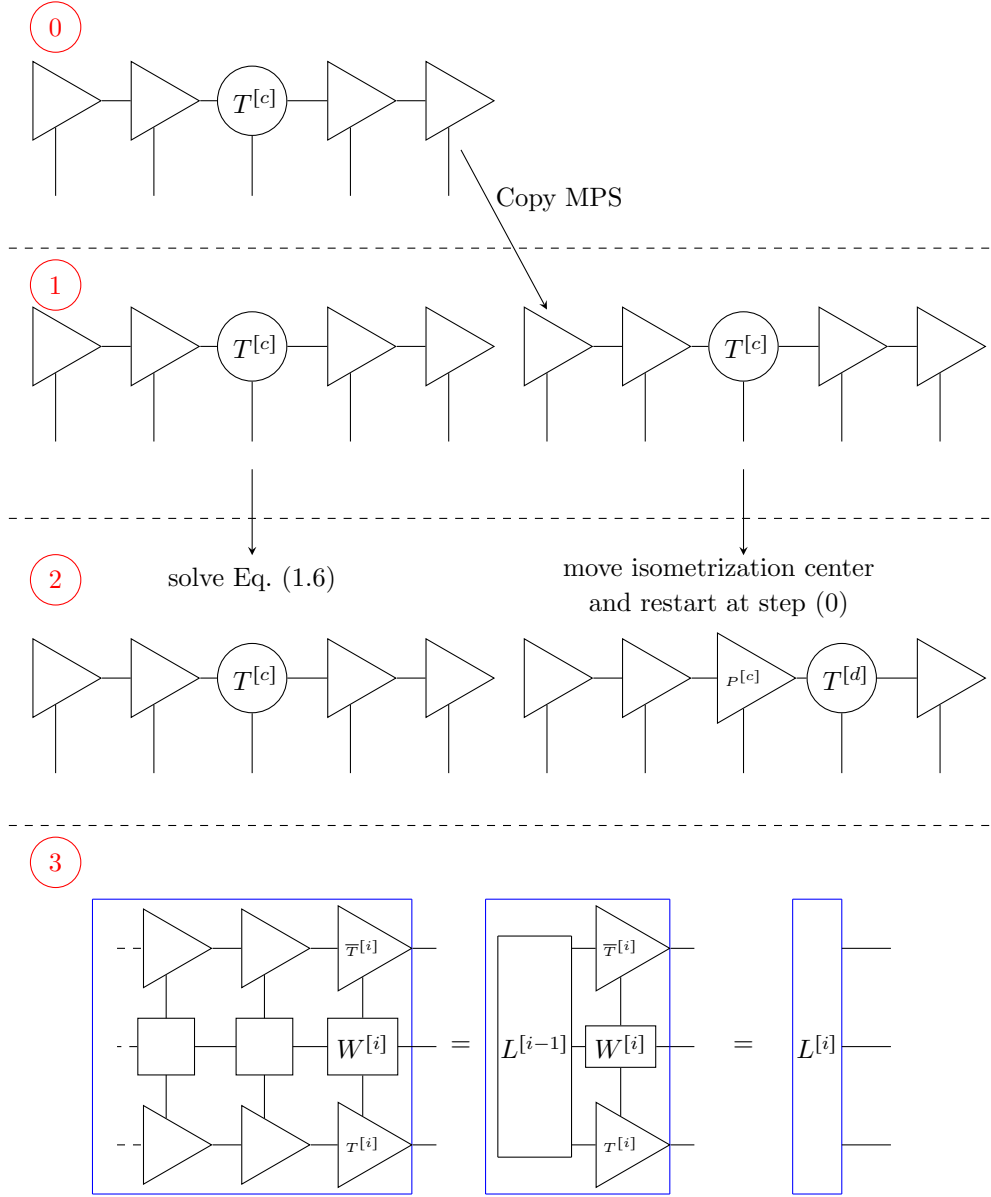
\begin{figure}
	\centering
	\begin{tikzpicture}[scale=0.65]
	\begin{pgfonlayer}{nodelayer}
		\node [style=NONE] (269) at (0, 13.5) {};
		\node [style=NONE] (270) at (2, 13.5) {};
		\node [style=NONE] (273) at (4, 13.5) {};
		\node [style=NONE] (274) at (6, 13.5) {};
		\node [style=NONE] (277) at (8, 13.5) {};
		\node [style=none] (295) at (8.25, 14.5) {};
		\node [style=none] (296) at (10, 11.25) {};
		\node [style=none] (297) at (10.25, 13.5) {Copy MPS};
		\node [style=none] (298) at (-1, 12.5) {};
		\node [style=none] (299) at (19, 12.5) {};
		\node [style=none] (310) at (4, 6.25) {};
		\node [style=none] (333) at (-1, 7) {};
		\node [style=none] (334) at (19, 7) {};
		\node [style=none] (335) at (14, 6.25) {};
		\node [style=none] (336) at (4, 8) {};
		\node [style=none] (337) at (14, 8) {};
		\node [style=none] (339) at (4, 5.75) {solve \autoref{eq:loc_diff_eq}};
		\node [style=none] (340) at (13.5, 6) {move isometrization center};
		\node [style=red text] (341) at (0, 17) {0};
		\node [style=red text] (342) at (0, 11.75) {1};
		\node [style=red text] (343) at (0, 5.75) {2};
		\node [style=none] (354) at (-1, 0.75) {};
		\node [style=none] (355) at (19, 0.75) {};
		\node [style=none] (377) at (1.25, -1.75) {};
		\node [style=none] (378) at (1.25, -5.75) {};
		\node [style=none] (382) at (1.25, -3.75) {};
		\node [style=none] (383) at (0.75, -0.75) {};
		\node [style=none] (384) at (0.75, -6.75) {};
		\node [style=none] (387) at (7.25, -6.75) {};
		\node [style=none] (388) at (7.25, -0.75) {};
		\node [style=none] (389) at (7.75, -1.75) {};
		\node [style=none] (390) at (7.75, -3.75) {};
		\node [style=none] (391) at (7.75, -5.75) {};
		\node [style=none] (399) at (17, -1.75) {};
		\node [style=none] (400) at (17, -3.75) {};
		\node [style=none] (401) at (17, -5.75) {};
		\node [style=none] (402) at (16, -0.75) {};
		\node [style=none] (403) at (16, -6.75) {};
		\node [style=none] (404) at (16, -3.75) {};
		\node [style=none] (405) at (15, -6.75) {};
		\node [style=none] (406) at (15, -0.75) {};
		\node [style=none] (407) at (16, -1.75) {};
		\node [style=none] (408) at (16, -5.75) {};
		\node [style=none] (409) at (15.5, -3.75) {$L^{[i]}$};
		\node [style=red text] (413) at (0, 0) {3};
		\node [style={R_triangle}] (414) at (2.25, -1.75) {};
		\node [style={R_triangle}] (415) at (4.25, -1.75) {};
		\node [style={R_triangle}] (416) at (2.25, -5.75) {};
		\node [style={R_triangle}] (417) at (4.25, -5.75) {};
		\node [style=plain circle] (423) at (4, 15.5) {$T^{[c]}$};
		\node [style={R_triangle}] (424) at (6, 15.5) {};
		\node [style={R_triangle}] (425) at (8, 15.5) {};
		\node [style={R_triangle}] (426) at (2, 15.5) {};
		\node [style={R_triangle}] (427) at (0, 15.5) {};
		\node [style=NONE] (428) at (0, 8.5) {};
		\node [style=NONE] (429) at (2, 8.5) {};
		\node [style=NONE] (430) at (4, 8.5) {};
		\node [style=NONE] (431) at (6, 8.5) {};
		\node [style=NONE] (432) at (8, 8.5) {};
		\node [style=plain circle] (433) at (4, 10.5) {$T^{[c]}$};
		\node [style={R_triangle}] (434) at (6, 10.5) {};
		\node [style={R_triangle}] (435) at (8, 10.5) {};
		\node [style={R_triangle}] (436) at (2, 10.5) {};
		\node [style={R_triangle}] (437) at (0, 10.5) {};
		\node [style=NONE] (438) at (10, 8.5) {};
		\node [style=NONE] (439) at (12, 8.5) {};
		\node [style=NONE] (440) at (14, 8.5) {};
		\node [style=NONE] (441) at (16, 8.5) {};
		\node [style=NONE] (442) at (18, 8.5) {};
		\node [style=plain circle] (443) at (14, 10.5) {$T^{[c]}$};
		\node [style={R_triangle}] (444) at (16, 10.5) {};
		\node [style={R_triangle}] (445) at (18, 10.5) {};
		\node [style={R_triangle}] (446) at (12, 10.5) {};
		\node [style={R_triangle}] (447) at (10, 10.5) {};
		\node [style=NONE] (448) at (0, 1.75) {};
		\node [style=NONE] (449) at (2, 1.75) {};
		\node [style=NONE] (450) at (4, 1.75) {};
		\node [style=NONE] (451) at (6, 1.75) {};
		\node [style=NONE] (452) at (8, 1.75) {};
		\node [style=plain circle] (453) at (4, 3.75) {$T^{[c]}$};
		\node [style={R_triangle}] (454) at (6, 3.75) {};
		\node [style={R_triangle}] (455) at (8, 3.75) {};
		\node [style={R_triangle}] (456) at (2, 3.75) {};
		\node [style={R_triangle}] (457) at (0, 3.75) {};
		\node [style=NONE] (458) at (10, 1.75) {};
		\node [style=NONE] (459) at (12, 1.75) {};
		\node [style=NONE] (460) at (14, 1.75) {};
		\node [style=NONE] (461) at (16, 1.75) {};
		\node [style=NONE] (462) at (18, 1.75) {};
		\node [style={R_triangle}] (465) at (18, 3.75) {};
		\node [style={R_triangle}] (466) at (12, 3.75) {};
		\node [style={R_triangle}] (467) at (10, 3.75) {};
		\node [style={R_triangle}] (468) at (14, 3.75) {\tiny$P^{[c]}$};
		\node [style=plain circle] (469) at (16, 3.75) {$T^{[d]}$};
		\node [style=none] (476) at (10.5, -1.5) {};
		\node [style=none] (477) at (10.5, -6) {};
		\node [style=none] (478) at (10.5, -3.75) {};
		\node [style=none] (479) at (9, -6) {};
		\node [style=none] (480) at (9, -1.5) {};
		\node [style=none] (481) at (10.5, -1.75) {};
		\node [style=none] (482) at (10.5, -5.75) {};
		\node [style=none] (483) at (9.75, -3.75) {$L^{[i-1]}$};
		\node [style=none] (491) at (13, -1.75) {};
		\node [style=none] (492) at (13, -3.75) {};
		\node [style=none] (493) at (13, -5.75) {};
		\node [style=small kernel tikzit] (496) at (11.5, -3.75) {$W^{[i]}$};
		\node [style={R_triangle}] (497) at (11.5, -1.75) {\tiny$\overline{T}^{[i]}$};
		\node [style={R_triangle}] (498) at (11.5, -5.75) {\tiny$T^{[i]}$};
		\node [style=medium kernel tikzit] (499) at (4.25, -3.75) {};
		\node [style=medium kernel tikzit] (500) at (6.25, -3.75) {$W^{[i]}$};
		\node [style=medium kernel tikzit] (501) at (2.25, -3.75) {};
		\node [style=none] (502) at (8.75, -0.75) {};
		\node [style=none] (503) at (8.75, -6.75) {};
		\node [style=none] (504) at (12.5, -6.75) {};
		\node [style=none] (505) at (12.5, -0.75) {};
		\node [style=none] (506) at (8.25, -3.75) {=};
		\node [style=none] (507) at (14, -3.75) {=};
		\node [style={R_triangle}] (508) at (6.25, -1.75) {\tiny$\overline{T}^{[i]}$};
		\node [style={R_triangle}] (509) at (6.25, -5.75) {\tiny$T^{[i]}$};
		\node [style=none] (511) at (13.5, 5.25) {and restart at step (0)};
	\end{pgfonlayer}
	\begin{pgfonlayer}{edgelayer}
		\draw [style=dashed edge] (299.center) to (298.center);
		\draw [style=dashed edge] (334.center) to (333.center);
		\draw [style=right arrow] (295.center) to (296.center);
		\draw [style=right arrow] (336.center) to (310.center);
		\draw [style=right arrow] (337.center) to (335.center);
		\draw [style=dashed edge] (355.center) to (354.center);
		\draw [style=blue] (383.center) to (388.center);
		\draw [style=blue] (388.center) to (387.center);
		\draw [style=blue] (387.center) to (384.center);
		\draw [style=blue] (384.center) to (383.center);
		\draw [style=blue] (406.center) to (405.center);
		\draw [style=blue] (405.center) to (403.center);
		\draw [style=blue] (403.center) to (402.center);
		\draw [style=blue] (402.center) to (406.center);
		\draw (408.center) to (401.center);
		\draw (400.center) to (404.center);
		\draw (407.center) to (399.center);
		\draw (414) to (416);
		\draw (415) to (417);
		\draw [style=dashed edge] (377.center) to (414);
		\draw [style=dashed edge] (378.center) to (416);
		\draw (425) to (277);
		\draw (274) to (424);
		\draw (423) to (273);
		\draw (426) to (270);
		\draw (269) to (427);
		\draw (435) to (432);
		\draw (431) to (434);
		\draw (433) to (430);
		\draw (436) to (429);
		\draw (428) to (437);
		\draw (445) to (442);
		\draw (441) to (444);
		\draw (443) to (440);
		\draw (446) to (439);
		\draw (438) to (447);
		\draw (455) to (452);
		\draw (451) to (454);
		\draw (453) to (450);
		\draw (456) to (449);
		\draw (448) to (457);
		\draw (465) to (462);
		\draw (466) to (459);
		\draw (458) to (467);
		\draw (468) to (460);
		\draw (461) to (469);
		\draw (416) to (417);
		\draw (415) to (414);
		\draw (427) to (426);
		\draw (426) to (423);
		\draw (423) to (424);
		\draw (424) to (425);
		\draw (437) to (436);
		\draw (436) to (433);
		\draw (433) to (434);
		\draw (434) to (435);
		\draw (447) to (446);
		\draw (446) to (443);
		\draw (443) to (444);
		\draw (444) to (445);
		\draw (457) to (456);
		\draw (456) to (453);
		\draw (453) to (454);
		\draw (454) to (455);
		\draw (467) to (466);
		\draw (466) to (468);
		\draw (468) to (469);
		\draw (469) to (465);
		\draw (496) to (492.center);
		\draw (478.center) to (496);
		\draw (481.center) to (497);
		\draw (497) to (491.center);
		\draw (498) to (493.center);
		\draw (498) to (482.center);
		\draw [style=dashed edge] (501) to (382.center);
		\draw (501) to (390.center);
		\draw (480.center) to (476.center);
		\draw (476.center) to (477.center);
		\draw (477.center) to (479.center);
		\draw (479.center) to (480.center);
		\draw [style=blue] (502.center) to (505.center);
		\draw [style=blue] (505.center) to (504.center);
		\draw [style=blue] (504.center) to (503.center);
		\draw [style=blue] (503.center) to (502.center);
		\draw (498) to (496);
		\draw (496) to (497);
		\draw (417) to (509);
		\draw (509) to (391.center);
		\draw (500) to (509);
		\draw (500) to (508);
		\draw (508) to (389.center);
		\draw (508) to (415);
	\end{pgfonlayer}
\end{tikzpicture}
\caption{\label{fig:Paralellized-update} \textbf{Step (0) $\rightarrow$ (1):} The MPS with the core-tensor as the isometrization-center is copied in memory. \textbf{Step (1) $\rightarrow$ (2):} The computationally demanding step of solving the tensor differential equation for $T^{[c]}$ is performed while moving the isometrization center on the copy. The copy is then updated analogously, restarting from step (0). \textbf{(3):} Obtaining the left environment tensor as in \autoref{eq:left_env_tensors}.}
\end{figure}

The left environment tensors are defined via
\begin{equation}\label{eq:left_env_tensors}
	\mathcal{L}^{[i]}_{v_iw_i\bar{v}_i} = \sum \mathcal{L}^{[i-1]}_{\bar{v}_{i-1}z_i\bar{v}_{i-1}} T_{v_{i-1}s_iv_i}^{[i]} W_{w_{i-1}s_iz_iw_i} \bar{T}^{[i]}_{\bar{v}_{i-1} z_i \bar{v}_{i}},
\end{equation}
with $\mathcal{L}^{[0]}=1$. The right environment tensors $\mathcal{R}^{[i]}$ are obtained analogously. It is advisable to store and reuse the environment tensors whenever possible to avoid unnecessary repeated contractions. Solving the differential equation \autoref{eq:loc_diff_eq} is generally the most time-consuming part of the algorithm. Thus, even though the canonical center must be moved, the actual updates at each site are mostly performed in parallel. The parallelized update scheme is exemplified in \autoref{fig:Paralellized-update}.

Notably, for all but the root site, we merely need to find an isometric tensor with the correct range towards the root tensor. Thus, if we do not want to dynamically increase the basis, we can perform the QR decomposition on the solution of \autoref{eq:loc_diff_eq} $T'^{[c]}$ to obtain the updated tensor. Otherwise, we can increase the bond dimension towards the root tensor by stacking the original $T^{[c]}$ and updated $T'^{[c]}$ tensors and performing the QR decomposition on the result, yielding the new isometric tensor $Q'^{[c]}$. To make the next tensor's bond dimension compatible with the extended bond dimension, we must project it into the higher-dimensional space by contracting the virtual leg with the matrix
\begin{equation}
	\mathcal{M}^{[c]}_{v_c \hat{v}_c} = \sum \mathcal{M}^{[i-1]}_{v_{i-1}\hat{v}_{i-1}} Q^{[c]}_{v_{c-1}s_cv_c} Q'^{[c]}_{\hat{v}_{c-1} s_c \hat{v_c}},
\end{equation}
if the new tensor is to the right of $c$. Otherwise, use an analogous term contracting the previous $\mathcal{M}$ into the right virtual legs, if the new tensor is to the left of $c$. Notably, this definition shows that the augmentation step happens serially from the outside to the inside of the MPS. This also implies that we can only stop the dynamical adaptation of the bond dimension once at a bond from each side, but we may not restart it for a bond closer to the root tensor.

To avoid an uncontrolled expansion of the bond dimensions in the MPS, one performs a truncation of the MPS, for example, using the recursive truncation scheme shown in \cite{Ceruti2023}.

\begin{remark}[Assuming non-updated tensors to be static] \label{Remark:Staticness-assumption}
In the above differential equation \autoref{eq:loc_diff_eq}, only the state tensor $T^{[c]}$ depends on the time when solving it. Notably, the environment tensors do not change during a time step. Since the iteration runs in parallel for all tensors, the evolution of each tensor is performed assuming all other tensors are static until the next time step.
\end{remark}

\section{The Hybrid BUG Algorithm}
\subsection{Initializing the Hybrid MPS}

\begin{figure}
	\centering
	\input{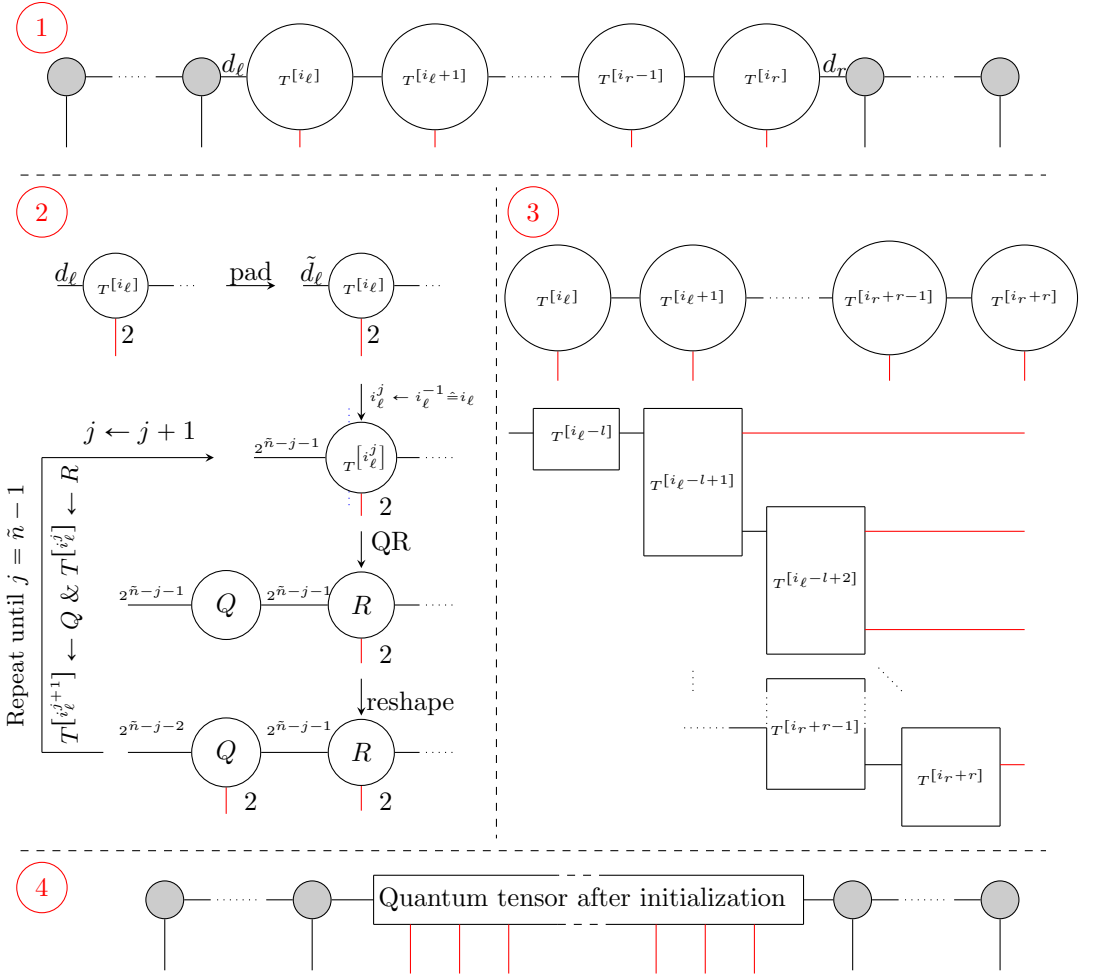}
\caption{Procedure for converting quantum tensors of an MPS into a quantum circuit. \textbf{(1):} The tensors of the MPS chosen to be part of the quantum tensor are shown in white, while the remaining tensors are gray. The physical legs that are part of the quantum tensor are drawn in red. \textbf{(2):} For the leftmost quantum tensor $T_{0}$, we pad the left bond until the bond dimension. Then the tensor is split via a QR-decomposition followed by a reshape of the resulting tensor $Q$ such that a new bond of dimension $2$ (red) is created. This leaves us with the initial situation on $Q$, and we iterate until the left bond-index is trivial. \textbf{(3):} The end product of the procedure applied to both sides, written in the manner of a quantum circuit of $T^{[i_l-l]}, \dots, T^{[i_r+r]}$.}
\label{fig:Procedure-for-converting}
\end{figure}

The initial assumption for the use of the proposed algorithm is that an execution instance of a purely classical time evolution on an MPS has run out of memory or is about to. 
At this point, we would decide to offload part of the MPS time evolution onto a quantum computer. For this, we have to convert the MPS into a hybrid tensor network by first designating parts of it as a quantum tensor.
\begin{definition}
	Let $\psi \in \mathbb{C}^{\prod_{i=1}^{r_c} d_{c,i} \prod_{i=1}^{r_q} d_{q,i}}$ be a \emph{quantum tensor} \cite{yuanQuantumSimulationHybrid2021} with quantum indices $j_{q,i} \in \{0, \dots, d_{q,i}\}$ Then, it is represented by the state of a physical quantum system. If the quantum tensor has additional classical indices $j_{c,i}$, then for each choice of values of the classical indices, a different state of a different physical quantum system is used to encode this part of the quantum tensor.
\end{definition}
In this work, we consider quantum tensors that have no classical indices and where the quantum state is a multi-qubit state, i.e., $r_c=0$ and $d_{q,i}=2$ for all $i$. Thus, the quantum tensor in question can be prepared via a common quantum circuit. Therefore, the task is to find the appropriate quantum circuit given the designated parts of the MPS. To this end, we assume that these parts are a consecutive chain of tensors, i.e., there are two sites $i_{\ell}$ and $i_{r}$ \label{intro-i-l} of the MPS that denote the leftmost and rightmost site that will be represented by the quantum tensor. 
(See \autoref{fig:Procedure-for-converting}, (1)). Note that if the tensors with the highest bond dimensions are not consecutive, it is possible to manipulate the MPS to make it so \cite{szalayTensorProductMethods2015}. Additionally, we must ensure that the isometrization center is on one of the tensors between and including $i_{\ell}$ and $i_{r}$.

When we designate part of the MPS as a quantum tensor, we must prepare it on a quantum computer by finding a corresponding circuit. However, existing procedures for encoding an MPS into a quantum state such as \cite{QskitcommunityMpstocircuit2025, ranEncodingMatrixProduct2020a} do not work on parts of an MPS, as the left bond dimension $d_\ell$ of the tensor $T^{[i_{\ell}]}$ and the right bond dimension $d_r$ of $T^{[i_{r}]}$ are not trivial. We can keep those bonds as classical bonds. This means, that we need one quantum state for every index-combination $(\ell, d_r) \in \{1, \dots, \ell\} \times \{1, \dots, d_r\}$. All of these would have to be evolved individually later on in the algorithm. We instead choose to trade off auxiliary qubits, such that only one quantum state has to be evolved. For this we use the procedure exemplified for tensor $T^{[i_{\ell}]}$ in \autoref{fig:Procedure-for-converting} (2). First we pad the left virtual bond of $T^{[i_{\ell}]}$ up to the dimension
\begin{equation}
	\tilde{d}_\ell = 2^{\tilde{n}} \quad \text{with } \tilde{n} = \text{argmin} \left( 2^n \geq d_\ell \right).
\end{equation}
Then we perform a QR decomposition separating the physical bond and right virtual bond of $T^{[i_{\ell}]}$ from the left virtual bond. The decomposition yields a new degree-3 tensor $R$ which we assign to $T^{[i_{\ell}]}$ and a matrix $Q \in \mathbb{C}^{\tilde{d}_\ell \times \tilde{d}_\ell}$. We can reshape the matrix into a new degree-3 tensor $T^{[0,\ell]} \in\mathbb{C}^{2^{\tilde{n}} \times 2 \times \tilde{d}_\ell}$, interpreting it as the new leftmost tensor of the sub MPS. We repeat this procedure until we find the leftmost tensor to be $T^{[i_{\ell}-\ell]}:=T^{[\tilde{n}-1,\ell]} \in\mathbb{C}^{1 \times 2 \times 2}$, i.e., the left virtual leg is trivial. We can perform the same procedure on the right of the sub-MPS. We denote the MPS resulting from this procedure as $\ket{\Psi_{\text{sub}}}$.

\label{explanation:reform-mps-into-circuit} Consider the re-written MPS $\ket{\Psi_{\text{sub}}}$, shown at (3) in \autoref{fig:Procedure-for-converting}. One may already spot a similarity to a quantum circuit, but there are still some required steps to find the appropriate quantum gates, as described in \cite{linRealImaginaryTimeEvolution2021}, for example.

Assume for ease of notation, that $T^{[k]}\in\mathbb{C}^{D_{k-1}\times2\times D_k}$ where $D_k=2^{\chi_k}$ for some $\chi_k \in \mathbb{N}$ for all $k$, otherwise we may pad the dimensions as explained before. Now one ensures that $\ket{\Psi_{\text{sub}}}$ is right-orthonormalized. We reinterpret $T^{[k]}$ for $i_{r}+r > k > i_{\ell}-\ell$ as a matrix
\begin{equation}
T^{[k]}\in\mathbb{C}^{D_{k-1}\times2\times D_k}\rightarrow \text{mat}\left(T^{[k]}\right)\in\mathbb{C}^{(2D_k) \times D_{k-1}}.
\end{equation}
Then the definition \autoref{eq:right_orthonormal} of right-canonical forms is equivalent to the rows of $\text{mat}\left(T^{[k]}\right)$ being orthonormal. Thus $\text{mat}\left(T^{[k]}\right)$ can be extend to a square unitary matrix  $U^{[k]}\in\mathbb{C}^{\tilde{D}\times\tilde{D}}$, where $\tilde{D} = \max (D_{k-1}, 2D_k)$. The situation is slightly different for the outermost tensors. $T^{[i_r+r]} \in \mathbb{C}^{2\times2}$ is already a unitary matrix and can be used without modification. The leftmost tensor $T^{[i_\ell-\ell]} \in \mathbb{C}^{2\times2}$ would be the isometrization center and thus not isometric. However, by introducing a trivial leg to the left, we can perform a QR decomposition and obtain an isometric tensor and a scalar. The scalar can be included in the bookkeeping matrices introduced in the next section. Then we can proceed as before by considering the tensor as $T^{[i_\ell-\ell]} \in \mathbb{C}^{1\times2\times2}$.

Using the unitary matrices $U^{[k]}$ as gates, we can implement $\ket{\Psi_{\text{sub}}}$ as quantum circuit
\begin{equation}
	\ket{\Psi_{\text{sub}}} \hat{=} \prod_{k=i_{\ell}-\ell}^{i_r+r} U^{[k]} \ket{0} = U_{\text{prep}} \ket{0} = \psi,
\end{equation}
where $\psi$ is the symbol used to refer to the quantum tensor from here on. Such a circuit for all $D_k = 2$ is shown in (3) in \autoref{fig:Procedure-for-converting}. We can see that the newly initialized input legs select the appropriate parts of the unitary matrices $U^{[k]}$ that correspond to the MPS tensors. The output legs correspond to the physical bond, and the virtual bond to the next tensor. Thus, the overall structure of the MPS remains evident.

Note that we chose this method because it leverages an intuitive connection between MPS and quantum circuits. However, there exist more advanced methods of MPS state-preparation \cite{rudolphDecompositionMatrixProduct2023, malzPreparationMatrixProduct2024, smithConstantdepthPreparationMatrix2024}.

\subsection{Updating the Quantum Tensor}

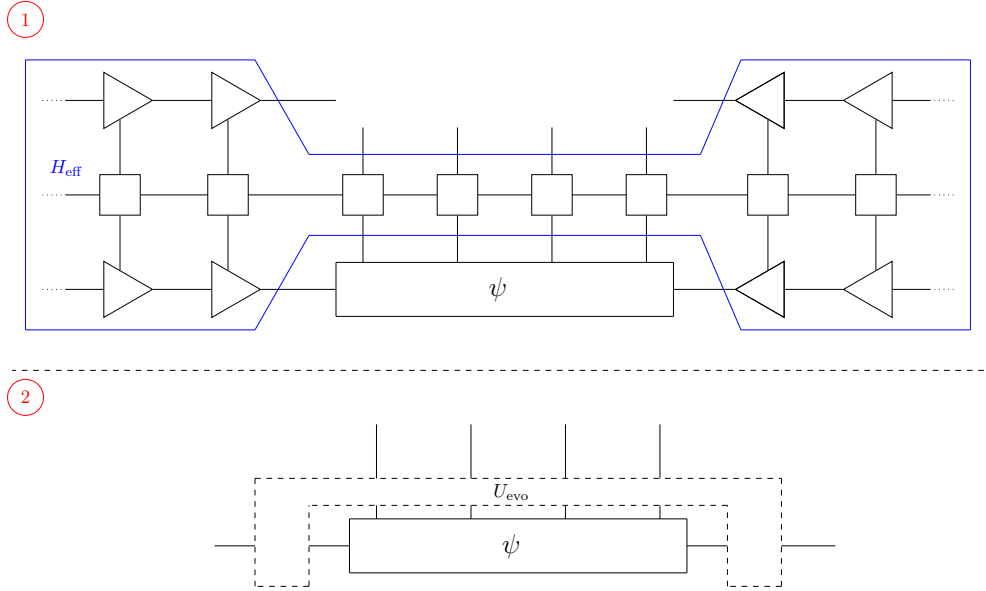
\begin{figure}
\centering
\resizebox{\textwidth}{!}{%
	\begin{tikzpicture}[]
	\begin{pgfonlayer}{nodelayer}
		\node [style={R_triangle}] (0) at (-6.5, -0.5) {};
		\node [style=NONE] (1) at (-8, -0.5) {};
		\node [style={R_triangle}] (2) at (-4.5, -0.5) {};
		\node [style=none] (7) at (-2.5, -0.5) {};
		\node [style=none] (8) at (3.75, -0.5) {};
		\node [style=none] (10) at (-2, -1) {};
		\node [style=none] (11) at (-0.25, -1) {};
		\node [style=none] (13) at (1.5, -1) {};
		\node [style=none] (14) at (3.25, -1) {};
		\node [style=none] (20) at (-7.5, -0.5) {};
		\node [style=none] (23) at (8.5, -0.5) {};
		\node [style=none] (24) at (9, -0.5) {};
		\node [style={L_triangle left}] (25) at (5.5, -0.5) {};
		\node [style={L_triangle left}] (26) at (7.5, -0.5) {};
		\node [style={L_triangle left}] (27) at (5.5, -0.5) {};
		\node [style={R_triangle}] (32) at (-6.5, -4) {};
		\node [style=NONE] (33) at (-8, -4) {};
		\node [style={R_triangle}] (34) at (-4.5, -4) {};
		\node [style=none] (35) at (-2.5, -3.5) {};
		\node [style=none] (36) at (-2.5, -4.5) {};
		\node [style=none] (37) at (3.75, -4.5) {};
		\node [style=none] (38) at (3.75, -3.5) {};
		\node [style=none] (39) at (-2.5, -4) {};
		\node [style=none] (40) at (3.75, -4) {};
		\node [style=none] (41) at (0.5, -4) {\Large$\psi$};
		\node [style=none] (42) at (-2, -3.5) {};
		\node [style=none] (43) at (-0.25, -3.5) {};
		\node [style=none] (44) at (1.5, -3.5) {};
		\node [style=none] (45) at (3.25, -3.5) {};
		\node [style=none] (46) at (-7.5, -4) {};
		\node [style=none] (47) at (8.5, -4) {};
		\node [style=none] (48) at (9, -4) {};
		\node [style={L_triangle left}] (49) at (5.5, -4) {};
		\node [style={L_triangle left}] (50) at (7.5, -4) {};
		\node [style={L_triangle left}] (51) at (5.5, -4) {};
		\node [style=medium kernel tikzit] (52) at (-6.5, -2.25) {};
		\node [style=medium kernel tikzit] (53) at (-4.5, -2.25) {};
		\node [style=medium kernel tikzit] (54) at (-2, -2.25) {};
		\node [style=medium kernel tikzit] (55) at (-0.25, -2.25) {};
		\node [style=medium kernel tikzit] (56) at (1.5, -2.25) {};
		\node [style=medium kernel tikzit] (57) at (3.25, -2.25) {};
		\node [style=medium kernel tikzit] (58) at (5.5, -2.25) {};
		\node [style=medium kernel tikzit] (59) at (7.5, -2.25) {};
		\node [style=none] (83) at (-2.25, -8.25) {};
		\node [style=none] (84) at (-2.25, -9.25) {};
		\node [style=none] (85) at (4, -9.25) {};
		\node [style=none] (86) at (4, -8.25) {};
		\node [style=none] (87) at (-2.25, -8.75) {};
		\node [style=none] (88) at (4, -8.75) {};
		\node [style=none] (89) at (0.75, -8.75) {\Large$\psi$};
		\node [style=none] (90) at (-1.75, -8.25) {};
		\node [style=none] (91) at (0, -8.25) {};
		\node [style=none] (92) at (1.75, -8.25) {};
		\node [style=none] (93) at (3.5, -8.25) {};
		\node [style=none] (123) at (-8.25, 0.25) {};
		\node [style=none] (124) at (-8.25, -4.75) {};
		\node [style=none] (125) at (9.25, -4.75) {};
		\node [style=none] (126) at (9.25, 0.25) {};
		\node [style=none] (127) at (-3, -3) {};
		\node [style=none] (128) at (-3, -1.5) {};
		\node [style=none] (129) at (4.25, -1.5) {};
		\node [style=none] (130) at (4.25, -3) {};
		\node [style=none] (131) at (-4, -4.75) {};
		\node [style=none] (132) at (-4, 0.25) {};
		\node [style=none] (133) at (5, -4.75) {};
		\node [style=none] (134) at (5, 0.25) {};
		\node [style=none] (136) at (-7.5, -1.75) {$\color{blue}{H_{\text{eff}}}$};
		\node [style=none] (137) at (9.5, -5.5) {};
		\node [style=none] (138) at (-8.5, -5.5) {};
		\node [style=red text] (139) at (-8.25, 1) {};
		\node [style=red text] (140) at (-8.25, 1) {1};
		\node [style=red text] (141) at (-8.25, -6) {};
		\node [style=red text] (142) at (-8.25, -6) {2};
		\node [style=none] (147) at (8.5, -2.25) {};
		\node [style=NONE] (148) at (9, -2.25) {};
		\node [style=none] (149) at (-7.5, -2.25) {};
		\node [style=NONE] (150) at (-8, -2.25) {};
		\node [style=none] (151) at (-1.75, -6.5) {};
		\node [style=none] (152) at (0, -6.5) {};
		\node [style=none] (153) at (1.75, -6.5) {};
		\node [style=none] (159) at (3.5, -6.5) {};
		\node [style=none] (160) at (-4, -7.5) {};
		\node [style=none] (161) at (-3, -8) {};
		\node [style=none] (162) at (4.75, -8) {};
		\node [style=none] (163) at (5.75, -7.5) {};
		\node [style=none] (164) at (0.75, -7.75) {$U_{\text{evo}}$};
		\node [style=none] (165) at (-1.75, -8) {};
		\node [style=none] (166) at (0, -8) {};
		\node [style=none] (167) at (1.75, -8) {};
		\node [style=none] (168) at (1.75, -7.5) {};
		\node [style=none] (169) at (0, -7.5) {};
		\node [style=none] (170) at (-1.75, -7.5) {};
		\node [style=none] (171) at (3.5, -7.5) {};
		\node [style=none] (172) at (3.5, -8) {};
		\node [style=none] (173) at (-3, -9.5) {};
		\node [style=none] (174) at (-4, -9.5) {};
		\node [style=none] (175) at (4.75, -9.5) {};
		\node [style=none] (176) at (5.75, -9.5) {};
		\node [style=none] (177) at (4.75, -8.75) {};
		\node [style=none] (178) at (6.75, -8.75) {};
		\node [style=none] (179) at (-3, -8.75) {};
		\node [style=none] (180) at (-4.75, -8.75) {};
		\node [style=none] (181) at (5.75, -8.75) {};
		\node [style=none] (182) at (-4, -8.75) {};
	\end{pgfonlayer}
	\begin{pgfonlayer}{edgelayer}
		\draw (0) to (2);
		\draw (2) to (7.center);
		\draw (0) to (20.center);
		\draw [style=dots] (20.center) to (1);
		\draw [style=dots] (23.center) to (24.center);
		\draw (8.center) to (27);
		\draw (27) to (26);
		\draw (26) to (23.center);
		\draw (32) to (34);
		\draw (34) to (39.center);
		\draw (35.center) to (36.center);
		\draw (36.center) to (37.center);
		\draw (37.center) to (38.center);
		\draw (38.center) to (35.center);
		\draw (32) to (46.center);
		\draw [style=dots] (46.center) to (33);
		\draw [style=dots] (47.center) to (48.center);
		\draw (40.center) to (51);
		\draw (51) to (50);
		\draw (50) to (47.center);
		\draw (32) to (52);
		\draw (53) to (34);
		\draw (42.center) to (54);
		\draw (55) to (43.center);
		\draw (44.center) to (56);
		\draw (57) to (45.center);
		\draw (51) to (58);
		\draw (59) to (50);
		\draw (0) to (52);
		\draw (2) to (53);
		\draw (10.center) to (54);
		\draw (11.center) to (55);
		\draw (13.center) to (56);
		\draw (14.center) to (57);
		\draw (27) to (58);
		\draw (26) to (59);
		\draw (83.center) to (84.center);
		\draw (84.center) to (85.center);
		\draw (85.center) to (86.center);
		\draw (86.center) to (83.center);
		\draw [style=blue] (123.center) to (124.center);
		\draw [style=blue] (123.center) to (132.center);
		\draw [style=blue] (132.center) to (128.center);
		\draw [style=blue] (128.center) to (129.center);
		\draw [style=blue] (129.center) to (134.center);
		\draw [style=blue] (134.center) to (126.center);
		\draw [style=blue] (126.center) to (125.center);
		\draw [style=blue] (125.center) to (133.center);
		\draw [style=blue] (133.center) to (130.center);
		\draw [style=blue] (130.center) to (127.center);
		\draw [style=blue] (127.center) to (131.center);
		\draw [style=blue] (131.center) to (124.center);
		\draw [style=dashed edge] (137.center) to (138.center);
		\draw (52) to (59);
		\draw (59) to (147.center);
		\draw (52) to (149.center);
		\draw [style=dots] (147.center) to (148);
		\draw [style=dots] (149.center) to (150);
		\draw (172.center) to (93.center);
		\draw (92.center) to (167.center);
		\draw (166.center) to (91.center);
		\draw (90.center) to (165.center);
		\draw [style=black] (151.center) to (170.center);
		\draw [style=black] (152.center) to (169.center);
		\draw [style=black] (153.center) to (168.center);
		\draw [style=black] (159.center) to (171.center);
		\draw [style=dashed edge] (173.center) to (161.center);
		\draw [style=dashed edge] (173.center) to (174.center);
		\draw [style=dashed edge] (160.center) to (174.center);
		\draw [style=dashed edge] (163.center) to (176.center);
		\draw [style=dashed edge] (176.center) to (175.center);
		\draw [style=dashed edge] (175.center) to (162.center);
		\draw [style=dashed edge] (162.center) to (161.center);
		\draw [style=dashed edge] (160.center) to (163.center);
		\draw [style=black] (88.center) to (177.center);
		\draw [style=black] (179.center) to (87.center);
		\draw [style=black] (180.center) to (182.center);
		\draw [style=black] (181.center) to (178.center);
	\end{pgfonlayer}
\end{tikzpicture}%
}
\caption{\label{fig:TDVP-update}\textbf{(1):} Showcasing the right-hand-side of the Schrödinger equation for the quantum tensor $\psi$ up the scalar factor $-i$. The effective Hamiltonian is enclosed by the blue line. \textbf{(2):} The actual update of the quantum tensor by applying the appropriate circuit $U_{\text{evo}}$.}
\end{figure}

Obtaining the environment tensors for each site is a preprocessing step that must be performed before each time step. Notably, this is also one of the few times where the quantum and classical tensors must communicate. We can construct all environment tensors $\mathcal{L}^{[i]}$ for $i<i_{\ell}$ and $\mathcal{R}^{[i]}$ for $i>i_{r}$ purely via classical contractions.

Due to the parallelism inherent in the BUG algorithm, we can update the quantum tensor without considering other parts of the MPS, except for the environment tensors. Similar to the purely classical BUG algorithm, we must solve a Schrödinger equation of the form \autoref{eq:loc_diff_eq} to update the quantum tensor $\psi$. The effective Hamiltonian is the contraction of the classical environment tensors $\mathcal{L}^{[i_\ell-1]}$ and $\mathcal{R}^{[i_r+1]}$ with the MPO tensors $W^{[i]}$ for $i \in \{i_\ell,\dots,i_r\}$ corresponding to the part of the MPS that was translated into the quantum tensor. This is the part outlined in blue in (1), \autoref{fig:TDVP-update}. As all the other tensors in the MPS are isometric, due to the canonical form, and the original MPO represents a Hermitian operator $H$, the effective Hamiltonian will also be Hermitian. Thus, to update the quantum tensor, we need to solve an actual Schrödinger equation. Doing this is one of the most promising tasks for quantum computers, with a variety of methods available, and one can, in principle, use any of them to obtain the new quantum state.

However, since we represent the quantum tensor as a quantum circuit, it is convenient to use a method that yields a quantum circuit $U_{\text{evo}}$ that evolves the quantum tensor according to $H_{\text{eff}}$. Then the updated quantum tensor is given by
\begin{equation}
	\psi(\Delta t) = U_{\text{evo}} \psi(0) = U_{\text{evo}} U_{\text{prep}} \ket{0},
\end{equation}
where $\Delta t$ is the time step size. To find $U_{\text{evo}}$, we recommend using the method proposed in \cite{Zhang2024}. It suffices to approximate the time evolution classically when applied to Haar-random product states. From the results, one can optimize an ansatz circuit to represent $U_{\text{evo}}$. While we cannot contract the full effective Hamiltonian, we can determine $H_{\text{eff}}^p \ket{\phi}$ as an MPS if $\ket{\phi}$ is a product state and $p$ is small enough to keep the bond dimensions low. This corresponds to a slightly modified application of an MPO to an MPS \cite{Camano2026, Milbradt2026}. Using the resulting MPS, we approximate $U_{\text{evo}} \ket{\phi}$ via the series expansion
\begin{equation}
    U_{\text{evo}} \ket{\phi} \approx \sum_{n=0}^{p} \frac{(-iHt)^n}{n!}.
\end{equation}

\subsection{Interaction between Classical and Quantum Tensors}

\begin{figure}
	\centering
	\resizebox{\textwidth}{!}{
		\input{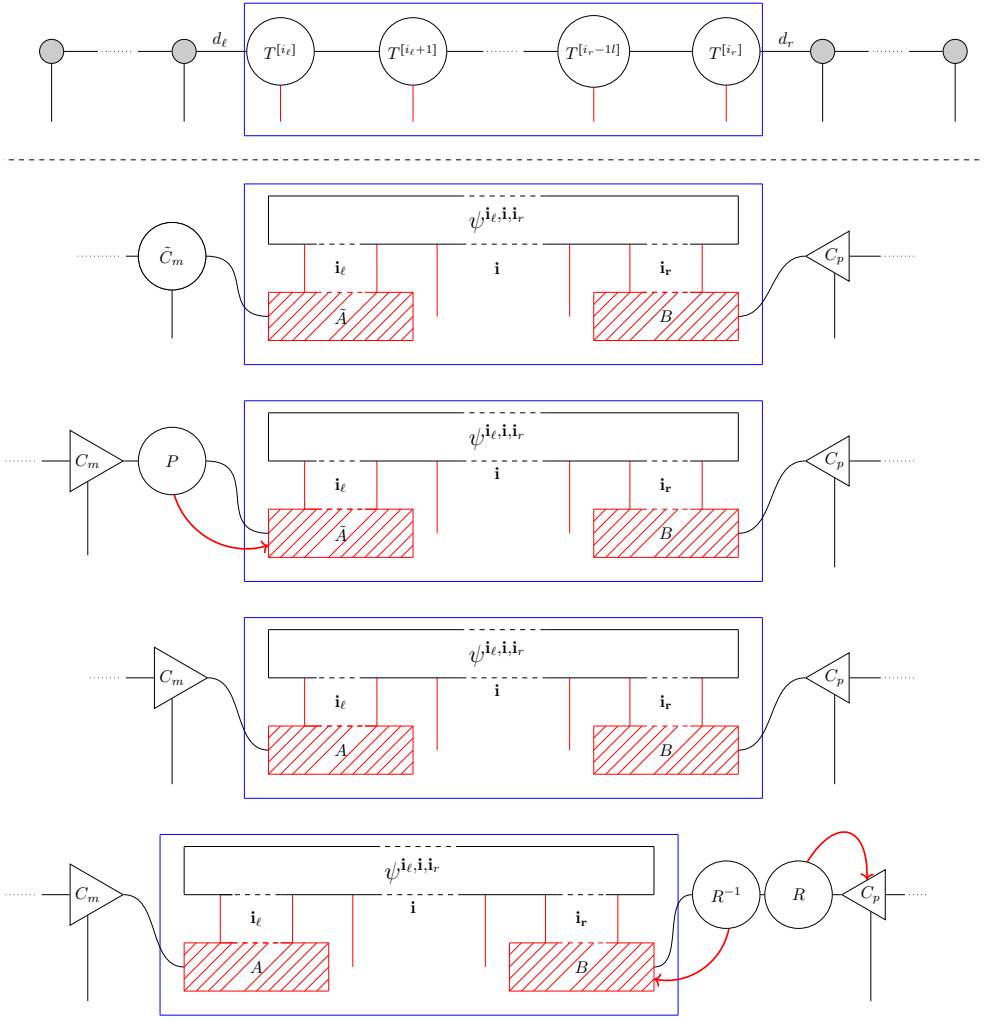}%
	}
\flushleft
	\caption{\label{fig:moving-isometry-center}\textbf{(Above, blue)} The classical part of the tensor network, which will be encoded in the quantum tensor $\psi$ and its bookkeeping matrices. \textbf{(Below, blue)} The effective quantum tensor including the classically stored bookkeeping matrices (red) and the quantum state $\psi$ which together encode the site tensors $T^{[i_l]}\dots T^{[i_r]}$. \newline \textbf{(Below)} \textit{Moving of the isometry center over the effective quantum tensor.}
	\underline{Step (1):} Reshape and QR decompose $\tilde{C}_{m}$ in order to left-isometrize the classical tensor, where $\tilde{C}_{m}=C_{m}P$ 
	\underline{Step (2):} Define $M_{k,j}:=\Big(A_{\mathbf{i}_{l},k}B_{\mathbf{i}_{r},j}\psi^{\mathbf{i}_{l},\mathbf{i},\mathbf{i}_{r}}(t)\Big)^{\dagger}\Big(A_{\mathbf{i}_{l},k}B_{\mathbf{i}_{r},j}\psi^{\mathbf{i}_{l},\mathbf{i},\mathbf{i}_{r}}(t)\Big)$	and contract $P$ with the left bookkeeping-matrix $\tilde{A}_{\mathbf{i}_{l},k}$ of the quantum tensor $\psi$ 
	\underline{Step (3):} $M$ is now a positive semi-definite Hermitian matrix and as such can be diagonalized as $M=U^{\dagger}DU$. Define $R=\sqrt{D}U$ and insert the identity by setting the right bookkeeping matrix to $R^{-1}$ and contract $C_{p}$ with $R$. Thus, we have successfully moved the isometrization center over the quantum tensor.}
\end{figure}

While the quantum tensor update is performed, we can execute the following preparation steps for the classical update in parallel. This is the main part involving communication between the quantum and classical tensors.

\begin{definition}
	Let $\mathbf{i}_{\ell},\mathbf{i},\mathbf{i}_{r}, \mathbf{j}_{1}, \mathbf{j}_{2}$ be multi-indices, $\psi_{\mathbf{j}_{1},\mathbf{j}_{2}}^{\mathbf{i}_{l},\mathbf{i},\mathbf{i}_{r}}$ be a quantum tensor and $A_{\mathbf{i}_{\ell},\mathbf{j}_{1},k},B_{\mathbf{i}_{r}, \mathbf{j}_{2},l}$ be classical tensors. 
	We define $A_{\mathbf{i}_{\ell},\mathbf{j}_{1},k}$, $B_{\mathbf{i}_{r}, \mathbf{j}_{2},l}$ as the \emph{bookkeeping tensors} of the quantum tensor $\psi_{\mathbf{\mathbf{j}}_{1},\mathbf{j}_{2}}^{\mathbf{\mathbf{i}_{l},\mathbf{i},\mathbf{i_{r}}}}$ and the triplet 
	\begin{equation}
		(A_{\mathbf{i}_{\ell},\mathbf{j}_{1},k},B_{\mathbf{i}_{r}, \mathbf{j}_{2},l},\psi_{\mathbf{j}_{1},\mathbf{j}_{2}}^{\mathbf{i}_{l},\mathbf{i},\mathbf{i}_{r}})
	\end{equation}
	as the \emph{effective quantum tensor}, which logically results from the contraction
	\begin{equation}
	A_{\mathbf{i}_{\ell},\mathbf{j}_{1},k}B_{\mathbf{i}_{r}, \mathbf{j}_{2},l}\psi_{\mathbf{j}_{1},\mathbf{j}_{2}}^{\mathbf{i}_{l},\mathbf{i},\mathbf{i}_{r}}.
	\end{equation}
\end{definition}

Just like \cite{schuhmacherHybridTreeTensor2024}, we use superscript modes for quantum legs and subscript modes for classical legs. Let $t_{0}$ be the starting time of the executed time-evolution and $t_{1}$ the first time-step. In the following, different objects needed during the execution of the algorithm will be distinguished by their time. 

Consider the classical site-tensors $\ensuremath{T_{l,i_{l}}^{[i_{\ell}]}},\ensuremath{T_{i_{l},i_{l+1}}^{[i_{\ell}+1]}},\dots,\ensuremath{T_{i_{r-1},r}^{[i_{r}]}}$ designated for offloading to the quantum computer. After initialization using the procedure in \autoref{fig:Procedure-for-converting}, this will correspond to the quantum tensor $\psi^{\mathbf{i}_{l},\mathbf{i},\mathbf{i}_{r}}(t_{0})$, where the sub-systems $\mathbf{i}_{l},\mathbf{i}_{r}$ correspond to the additionally introduced qubits, representing the virtual bond dimension. Let the effective quantum tensor corresponding to $\psi^{\mathbf{i}_{l},\mathbf{i},\mathbf{i}_{r}}(t_{0})$ be $\Big(A_{\mathbf{i}_{l},k},B_{\mathbf{i}_{r},j},\psi^{\mathbf{i}_{l},\mathbf{i},\mathbf{i}_{r}}\Big)$. Given that $\psi^{\mathbf{i}_{l}}$ is the isometrization-centre, we can initialise the two bookkeeping tensors as the identity. Using the original MPS-decomposition of the quantum tensor, the form of the bookkeping tensors for the left- and right-isometrized quantum tensor can be determined without the use of quantum resources.\\

Once we move to a later time step, this computation is non-trivial. For this we introduce the matrix
\begin{equation}
    M_{k,j}(t):=\Big(A_{\mathbf{i}_{l},k}B_{\mathbf{i}_{r},z}\psi^{\mathbf{i}_{l},\mathbf{i},\mathbf{i}_{r}}(t)\Big)^{\dagger}\Big(A_{\mathbf{i}_{l},j}B_{\mathbf{i}_{r},z}\psi^{\mathbf{i}_{l},\mathbf{i},\mathbf{i}_{r}}(t)\Big),
\end{equation}
which is a positive semi-definite Hermitian matrix \cite{schuhmacherHybridTreeTensor2024}. Thus, it can be diagonalized as $M_{k,j}(t)=U^{\dagger}DU$. Then we insert the identity between the right bookkeeping matrix and the classical tensor directly to the right of the effective quantum tensor as $\mathbb{1}=R^{-1}R$ with $R:=\sqrt{D}U$. This is shown in \autoref{fig:moving-isometry-center}. Then, $R^{-1}$ can be contracted with the right bookkeeping matrix $B_{\mathbf{i}_{r},j}$ and $R$ with the classical site-tensor $c_{P}$ directly to the right of the effective quantum tensor. This results in the isometrization center now being at the site tensor, which was previously $c_{P}$ (see also \autoref{fig:moving-isometry-center}).

For this operation to be executed, $M_{k,j}(t)$ needs to be determined. Consider that executing a time-step corresponds to applying a unitary circuit $U_{\text{evo}}$ to the full quantum tensor, i.e.
\begin{align*}
M_{k,j}(t) & =\Big((A_{\mathbf{i}_{l},k}\otimes\mathbb{1}_{\mathbf{i}}\otimes B_{\mathbf{i}_{r},z})U_{\text{evo}}\psi^{\mathbf{i}_{l},\mathbf{i},\mathbf{i}_{r}}(t_{0})\Big)^{\dagger}\Big((A_{\mathbf{i}_{l},j}\otimes\mathbb{1}_{\mathbf{i}}\otimes B_{\mathbf{i}_{r},z})U_{\text{evo}}\psi^{\mathbf{i}_{l},\mathbf{i},\mathbf{i}_{r}}(t_{0})\Big)\\
 & =\Big((A_{\mathbf{i}_{l},k}\otimes\mathbb{1}_{\mathbf{i}}\otimes B_{\mathbf{i}_{r},z})\psi^{\mathbf{i}_{l},\mathbf{i},\mathbf{i}_{r}}(t_{1})\Big)^{\dagger}\Big((A_{\mathbf{i}_{l},j}\otimes\mathbb{1}_{\mathbf{i}}\otimes B_{\mathbf{i}_{r},z})\psi^{\mathbf{i}_{l},\mathbf{i},\mathbf{i}_{r}}(t_{1})\Big).
\end{align*}
This corresponds to only determining expectation values involving subsystems $\mathbf{i}_{l}$ and $\mathbf{i}_{r}$, with size determined by the number of auxiliary qubits. Since this number exhibits favorable scaling behavior (the number of auxiliary qubits scales with $O(\log(n))$ of the bond dimension of the bond between the classical and the quantum tensor before application of \autoref{fig:Procedure-for-converting}), this operation is not too demanding in terms of quantum resources. 

For right-isometrization, analogous considerations holds: Once, before initializing the quantum tensor, we need to compute $M_{k,j}$ such that $\Big(A_{\mathbf{i}_{l},k}, B_{\mathbf{i}_{r},j}, \psi^{\mathbf{i}_{l},\mathbf{i},\mathbf{i}_{r}}(t_{0})\Big)$ is right-isometrized and then, updates can be done with a well-scaling workload for the quantum-device. 

Once the isometrization center is moved, we can determine the missing environment tensors. To update the classical tensors $T^{[i_{\ell}-1]}$ and $T^{[i_{\ell}+1]}$ we must first obtain $\mathcal{R}^{[i_{\ell}]}$ and $\mathcal{L}^{[i_{r}]}$, respectively. These immediately involve the full quantum tensor. For the recursive contraction, we must find the expectation value of the sub-MPO elements
\begin{align}
	h_{k,\ell} = \sum W^{[i_{\ell}]}_{ks_{i_{\ell}-1}z_{i_{\ell}}w_{i_{\ell}}} \cdots W^{[i_{r}]}_{ks_{i_{r}-1}z_{i_{r}}w_{i_{r}}} \cdot \ket{z_{i_{\ell}}, \dots, z_{i_{r}}}\bra{s_{i_{\ell}}, \dots, s_{i_{r}}}
\end{align}
enlarged by chains containing Pauli $X$ and $Z$ operators at both ends, to account for the qubits representing the virtual bond indices. While this evaluation seems expensive, there are non-trivial methods that can improve the performance of this substep \cite{Kohda2022, Elben2023}. These methods work well in our framework, as it is trivial to determine expectation values of the partial Hamiltonian with regard to product states in the computational basis. As soon as the environment tensors involving the quantum tensor are found, they can be reused to obtain the remaining environment tensors. Additionally, this costly process must only be performed once per time step.

\subsection{Parallel Update of Classical Tensors}
Once the environment tensors are determined, we can update the classical MPS tensors of the left and right chain as discussed in \autoref{sec:class_bug}. The quantum tensors would concern us only during the steps where the bond dimensions are varied. That is, during the augmentation and the eventual truncation of the virtual bonds.

However, we will stop the augmentation when reaching the bond towards the quantum tensor. This is a reasonable assumption, as quantum resources are primarily limited by qubit counts. Thus, most available qubits would already be in use. Additionally, the padding of the dimension during initialization to the nearest power of two could be increased to the next nearest power instead. Due to the exponential scaling behavior, this should allow for a sufficient bond dimension over the time step, since the BUG algorithm increases the bond dimension at most by a factor of $2$ in any case.

As we do not increase the bond dimensions between classical and quantum tensors, we do not need to truncate the corresponding bonds. Additionally, the recursive truncation procedure introduced in \cite{Ceruti2023} allows us to truncate the left and right sub MPSs without considering the quantum tensor between them. The fact that we can perform these post-processing steps independently of the quantum tensor allows us to further improve the parallelism of our proposed method.

During execution of the parallel BUG, the main iteration is running over every tensor in parallel, as remarked in \autoref{Remark:Staticness-assumption}. This allows our algorithm to update the classical and quantum tensors in parallel for each time step $\Delta t$. However, the classical update process on the classical tensors is likely faster than the quantum tensor update. Therefore, to avoid any idle time of the classical resources, we could evolve the classical tensors by multiple smaller time steps $\frac{\Delta t}{N}$, making it more exact. The interaction of the classical and quantum tensors would then occur whenever the times coincide, thereby reducing latency issues in the HPCQC integration.

\subsection{A Bird's Eye View}
\begin{figure}
\centering
\resizebox{\textwidth}{!}{%
	\begin{tikzpicture}
	\begin{pgfonlayer}{nodelayer}
		\node [style=process] (0) at (-10, 4) {Quantum tensor $\psi$ from sub-MPS};
		\node [style=process] (1) at (-10, 2) {Copy $\psi_\ell$ \& $\psi_r$};
		\node [style=return value] (2) at (-6.5, 2.5) {$\psi_r$};
		\node [style=return value] (3) at (-6.5, 1.25) {$\psi_\ell$};
		\node [style=return value] (4) at (-10, 0.75) {$\psi$};
		\node [style=none] (7) at (-2.5, 2.5) {};
		\node [style=process] (8) at (3, 4) {Get left $\ket{\Psi_\ell}$ and right $\ket{\Psi_r}$ sub-MPS};
		\node [style=process] (10) at (3, 1.25) {Classical Contraction to obtain environment tensors $\left\{\mathcal{L}^{[j]}\right\}_{j=0}^{i_\ell-1}$ and $\left\{\mathcal{R}^{[j]}\right\}_{j=i_r+1}^{L-1}$};
		\node [style=return value] (11) at (1.5, -1.5) {$\mathcal{R}^{[i_r+1]}$};
		\node [style=return value] (12) at (4.25, -1.5) {$\mathcal{L}^{[i_\ell-1]}$};
		\node [style=return value] (13) at (8, 0.25) {$\left\{\mathcal{L}^{[j]}\right\}_{j=0}^{i_\ell-2}$, $\ket{\Psi_\ell}$};
		\node [style=return value] (14) at (8.25, 2) {$\left\{\mathcal{R}^{[j]}\right\}_{j=i_r+2}^{L-1}$, $\ket{\Psi_r}$};
		\node [style=process] (15) at (1.5, -2.5) {};
		\node [style=process] (16) at (1.5, -2.5) {Copy};
		\node [style=process] (17) at (4.25, -3.25) {Copy};
		\node [style=none] (20) at (-5.5, -2.5) {};
		\node [style=none] (21) at (-1.5, -3.25) {};
		\node [style=return value] (22) at (2.75, -4.75) {};
		\node [style=none] (23) at (4.25, -4.75) {};
		\node [style=none] (24) at (1.5, -4.75) {};
		\node [style=process] (25) at (-10, -5.25) {Update $\psi$};
		\node [style=none] (26) at (2.75, -5.25) {};
		\node [style=process] (27) at (-6.5, -7.5) {Move iso. centre to site $i_\ell$};
		\node [style=process] (28) at (2, -7.5) {Move iso. centre to site $i_r$};
		\node [style=none] (29) at (-2.5, -6.75) {};
		\node [style=none] (30) at (2, -6.75) {};
		\node [style=none] (31) at (8, -5.5) {};
		\node [style=none] (32) at (-5, -5.5) {};
		\node [style=none] (33) at (10.75, 2) {};
		\node [style=none] (34) at (10.75, -6) {};
		\node [style=none] (35) at (3, -6) {};
		\node [style=process] (36) at (-6.5, -9) {Update $\ket{\Psi_\ell}$};
		\node [style=process] (37) at (2, -9) {Update $\ket{\Psi_rl}$};
		\node [style=decission] (38) at (2, -11.25) {Final time?};
		\node [style=decission] (39) at (-6.5, -11.25) {Final time?};
		\node [style=decission] (40) at (-10, -11.25) {Final time?};
		\node [style=none] (41) at (-2, -11.25) {};
		\node [style=none] (42) at (-2, -10) {};
		\node [style=return value] (43) at (5.75, -10) {};
		\node [style=none] (44) at (5.75, -11.25) {};
		\node [style=none] (45) at (5.75, 4) {};
		\node [style=none] (46) at (-12, -11.25) {};
		\node [style=none] (47) at (-12, 2) {};
		\node [style=return value] (48) at (-4, -14.5) {End};
		\node [style=none] (49) at (-10, -13.5) {};
		\node [style=none] (50) at (-6.5, -12.75) {};
		\node [style=none] (51) at (2, -13.25) {};
		\node [style=none] (52) at (-4, -12.75) {};
		\node [style=none] (53) at (-3.25, -13.25) {};
		\node [style=none] (54) at (-5, -13.5) {};
		\node [style=none] (55) at (2.5, -12.5) {True};
		\node [style=none] (56) at (-6, -12.5) {True};
		\node [style=none] (57) at (-9.5, -12.5) {True};
		\node [style=none] (58) at (4, -11) {False};
		\node [style=none] (59) at (-5, -11) {False};
		\node [style=none] (60) at (-11.5, -11) {False};
		\node [style=quantum process] (62) at (-6.5, -0.75) {Move iso. centre to bond $(i_\ell-1,i_\ell)$};
		\node [style=quantum process] (64) at (-2.5, -0.75) {Move iso. centre to bond $(i_r,i_r+1l)$};
		\node [style=quantum process] (66) at (-6.5, -4.25) {Obtain $\mathcal{R}^{[i_\ell]}$};
		\node [style=quantum process] (67) at (-2.5, -4.25) {Obtain $\mathcal{L}^{[i_r]}$};
	\end{pgfonlayer}
	\begin{pgfonlayer}{edgelayer}
		\draw [style=process connection] (1) to (4);
		\draw [style=process connection] (0) to (1);
		\draw [style=process connection] (1) to (2);
		\draw [style=process connection] (1) to (3);
		\draw [style=process no arrow] (2) to (7.center);
		\draw [style=process connection] (8) to (10);
		\draw [style=process connection] (10) to (11);
		\draw [style=process connection] (10) to (12);
		\draw [style=process connection] (10) to (13);
		\draw [style=process connection] (10) to (14);
		\draw [style=process connection] (11) to (16);
		\draw [style=process connection] (12) to (17);
		\draw [style=process no arrow] (21.center) to (17);
		\draw [style=process no arrow] (16) to (20.center);
		\draw [style=process no arrow] (16) to (24.center);
		\draw [style=process no arrow] (24.center) to (22);
		\draw [style=process no arrow] (23.center) to (22);
		\draw [style=process no arrow] (17) to (23.center);
		\draw [style=process connection] (4) to (25);
		\draw [style=process no arrow] (22) to (26.center);
		\draw [style=process connection] (26.center) to (25);
		\draw [style=process connection] (30.center) to (28);
		\draw [style=process no arrow] (29.center) to (30.center);
		\draw [style=process no arrow] (13) to (31.center);
		\draw [style=process no arrow] (31.center) to (32.center);
		\draw [style=process connection] (32.center) to (27);
		\draw [style=process no arrow] (33.center) to (34.center);
		\draw [style=process no arrow] (34.center) to (35.center);
		\draw [style=process connection] (35.center) to (28);
		\draw [style=process no arrow] (14) to (33.center);
		\draw [style=process connection] (27) to (36);
		\draw [style=process connection] (28) to (37);
		\draw [style=process connection] (37) to (38);
		\draw [style=process connection] (36) to (39);
		\draw [style=process connection] (25) to (40);
		\draw [style=process no arrow] (39) to (41.center);
		\draw [style=process no arrow] (41.center) to (42.center);
		\draw [style=process no arrow] (42.center) to (43);
		\draw [style=process no arrow] (38) to (44.center);
		\draw [style=process no arrow] (44.center) to (43);
		\draw [style=process no arrow] (43) to (45.center);
		\draw [style=process connection] (45.center) to (8);
		\draw [style=process no arrow] (40) to (46.center);
		\draw [style=process no arrow] (46.center) to (47.center);
		\draw [style=process connection] (47.center) to (1);
		\draw [style=process no arrow] (40) to (49.center);
		\draw [style=process no arrow] (49.center) to (54.center);
		\draw [style=process no arrow] (39) to (50.center);
		\draw [style=process no arrow] (50.center) to (52.center);
		\draw [style=process no arrow] (38) to (51.center);
		\draw [style=process no arrow] (51.center) to (53.center);
		\draw [style=process connection] (52.center) to (48);
		\draw [style=process connection] (54.center) to (48);
		\draw [style=process connection] (53.center) to (48);
		\draw [style=process connection] (3) to (62);
		\draw [style=process connection] (7.center) to (64);
		\draw [style=process connection] (62) to (66);
		\draw [style=process connection] (66) to (27);
		\draw [style=process connection] (20.center) to (66);
		\draw [style=process connection] (64) to (67);
		\draw [style=process connection] (21.center) to (67);
		\draw [style=process no arrow] (67) to (29.center);
	\end{pgfonlayer}
\end{tikzpicture}%
}
\caption{Flow chart of the main time-stepping update. The processes in the light grey dashed boxes require quantum resources.}
\label{fig:birds_eye}
\end{figure}
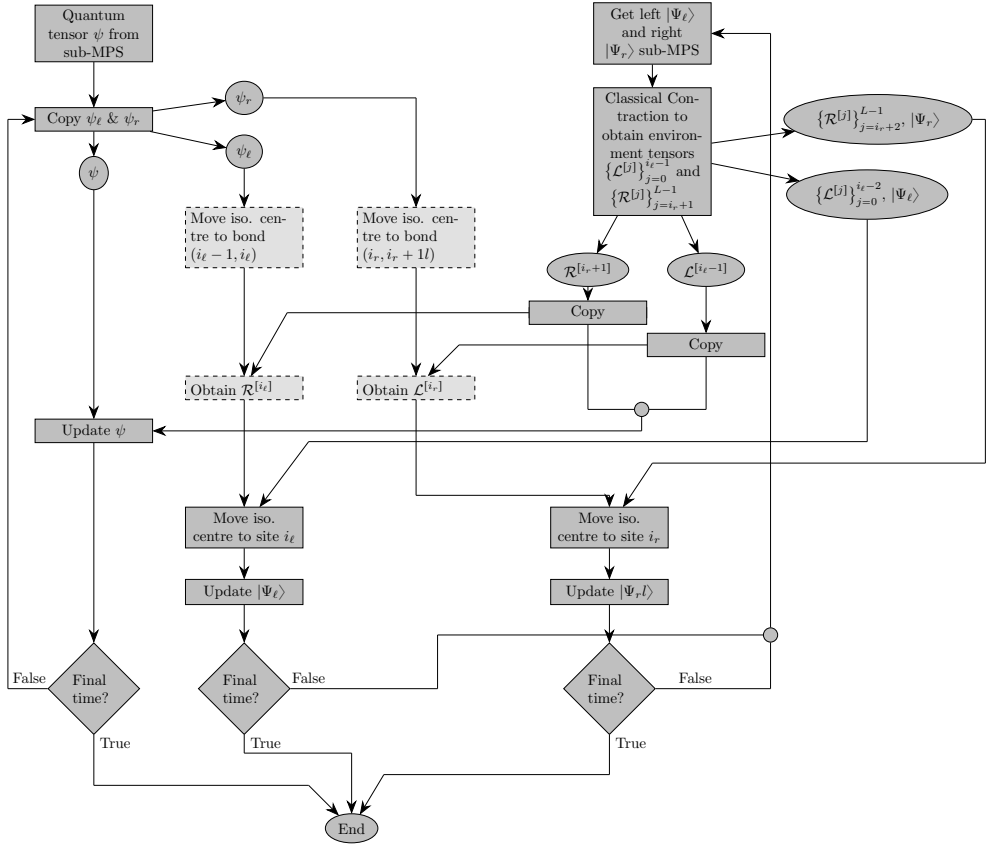
In this section, we present a concise summary of our algorithm, as illustrated in \autoref{fig:birds_eye}. We start with the assumption that we have a quantum tensor $\psi$ that represents the central part of the MPS and two chains of tensors, $\ket{\Psi_\ell}$ and $\ket{\Psi_r}$, that represent the left and right sub-MPS not contained in $\psi$. We must then compute the environment tensors that require only classical contractions $\left\{\mathcal{L}^{[j]}\right\}_{j=0}^{i_\ell-1}$ and $\left\{\mathcal{R}^{[j]}\right\}_{j=i_r+1}^{L-1}$. At this point, we can already create two copies $\psi_\ell$ and $\psi_r$ of the quantum tensor on which we move the isometrisation center to the bonds $(i_\ell -1, i_\ell)$ and $(i_r,i_r +1)$, respectively. This must occur on a quantum computer each. Once $\mathcal{L}^{[i_\ell-1]}$ and $\mathcal{R}^{[i_r+1]}$ are computed, the quantum tensor $\psi$ can be updated. In parallel, we compute the environment tensors $\mathcal{L}^{[i_r]}$ and $\mathcal{R}^{[i_\ell]}$ which is done using quantum resources. Once they are obtained, we can update the classical sub-MPS tensors, still in parallel with the update of the quantum tensor. At this point, we can update the classical tensors also with a smaller time step, if desired. Once all tensors are updated, we can repeat the procedure for the next time step.

\section{Conclusion}
To address the challenge of increasing memory footprint due to growing entanglement when simulating time evolution on classical computers, we have proposed a novel HPCQC algorithm based on the BUG algorithm. 
It has several notable features not present in previous works in the hTN framework, one of which is the parallelizability of classical and quantum parts of the algorithm. 
Furthermore, it allows a flexible trade-off between the number of quantum states constituting the quantum tensor and the number of qubits, down to evolving only one quantum state. 
Additionally, our algorithm is agnostic to which quantum time-evolution algorithm is used as a subroutine. Also, it can flexibly run a time evolution on classical computers until the classical memory is exhausted.

To enable practical implementation and benchmarking, we have described and explained the various components, parameters, and sub-processes of our algorithm in detail. 
Firstly, the conversion of a part of a tensor train into an MPS, and the conversion of said MPS into a quantum state. 
We have described how the isometrization-center is moved onto and over the quantum tensor. 
Furthermore, we described the generation of environment tensors for the hybrid tensor network and developed a definition of an effective quantum tensor. This is a further contribution to the theoretical framework of hybrid tensor networks laid down by \cite{yuanQuantumSimulationHybrid2021, schuhmacherHybridTreeTensor2024}. It clarifies how exactly a hybrid tensor train is encoded on a compute environment containing both quantum and classical computers. 

All in all, we have presented a parallel hybrid HPCQC algorithm designed for the EFTQC era and beyond. Even after the EFTQC era, we expect the possibilities of our presented technique of trading off qubits against the number of states to evolve to have an impact. This is because it showcases an algorithm that can use several quantum computers as accelerators for a classical compute cluster. 

\section{Discussion and Outlook}

The algorithm presented in this work lends itself to implementation in a hybrid HPCQC compute environment. 
Future work will focus on extensive numerical benchmarking, starting with spin lattice systems and moving on to small molecular systems. 
As the proposed algorithm includes several parameters that affect load balancing between HPC and QC devices, a multivariate evaluation should be conducted. This includes evaluating the balance among the number of auxiliary qubits, the number of states to evolve, and the number of quantum devices to use for time evolution, should several states be evolved. Furthermore, the time step sizes for the classical update steps and the quantum time evolution should be carefully matched. 
Another parameter is the choice of time-evolution algorithm and circuit synthesis method, as well as their hardware specificity. To this end, we consider Riemannian optimization, as described in \cite{leRiemannianQuantumCircuit2025}, for circuit-compression-paired methods like \cite{Zhang2024} as a suitable avenue for investigation. 
This work was presented using a tensor train framework for ease of understanding. Naturally, a further goal should be extension towards general tree tensor networks. 
Furthermore, practical implementation and benchmarking can help to give a detailed answer, if during the parallel update steps, the whole tensor is copied once for every sub tree as suggested by \cite{Ceruti2023} and then its isometrization-center is moved, or if the update steps are spawned during sequential moving of the isometrization center. 
Lastly, we leave it open to future work, how rank-adaptivity can be achieved for the quantum tensor using quantum algorithms. 

\section*{Acknowledgements}
The authors acknowledge the use of Gemini 3 (Google, version used: 28.03.2026) and Grammarly to polish the written text for grammar and general style.

The research by Tobias Valentin Bauer is partially and the research by Richard M. Milbradt is fully funded in the context of the Munich Quantum Valley, which is supported by the Bavarian state government with funds from the Hightech Agenda Bayern Plus. The research by TVB is furthermore funded by the German Federal Ministry of Research, Technology and Space (BMFTR) with the grants 13N16078 (MUNIQC-Atoms) and 13N16187 (MUNIQC-SC).

\section*{Authors' Contribution}
TVB and RMM conceptualised the projects and performed the derivation of the algorithms. In addition, they wrote the original draft. CBM supported during the derivation and writing process via discussion and editing.

\bibliographystyle{siamplain}
\bibliography{bibliothek}
\end{document}